\documentclass[11pt,a4paper]{article}

\usepackage[T1]{fontenc}
\usepackage[utf8]{inputenc}
\usepackage{lmodern}

\usepackage[margin=1in]{geometry}
\usepackage{setspace}
\usepackage{amsmath,amssymb}
\usepackage{siunitx}
\usepackage{booktabs}

\usepackage{graphicx}
\usepackage{caption}
\usepackage{float}

\usepackage{xcolor}
\usepackage{microtype}
\usepackage[hidelinks]{hyperref}

\title{Morphology Engineering of Mixed Ionic--Electronic Conductors through Aqueous Phase Separation}

\author{%
  Siqi Wang\textsuperscript{1,$\dagger$}%
  \and
  Maria A. Restrepo\textsuperscript{1,$\dagger$}%
  \and
  John Linkhorst\textsuperscript{2}%
  \and
  Matthias Wessling\textsuperscript{1,3,*}%
}

\date{}

\begin{document}
\maketitle

\begin{center}
\small
\textsuperscript{1}Chemical Process Engineering (AVT.CVT), RWTH Aachen University,\\
Forckenbeckstra\ss{}e 51, 52074 Aachen, Germany\\[0.4em]
\textsuperscript{2}Technische Universit\"at Darmstadt, Verfahrenstechnik elektrochemischer Systeme,\\
Otto-Berndt-Stra\ss{}e 2, 64287 Darmstadt, Germany\\[0.4em]
\textsuperscript{3}DWI -- Leibniz Institute for Interactive Materials e.V.,\\
Forckenbeckstra\ss{}e 50, 52074 Aachen, Germany\\[0.6em]
\textsuperscript{*}Correspondence: \texttt{manuscripts.cvt@avt.rwth-aachen.de}\\[0.4em]
$\dagger$~Siqi Wang and Maria A. Restrepo contributed equally to this publication.
\end{center}

\begin{abstract}
The performance of organic electrochemical transistors (OECTs) is fundamentally governed by the interplay between ionic accessibility and electronic transport within organic mixed ionic--electronic conductors. Although increasing channel thickness enhances transconductance, it also prolongs ion transport, resulting in the well-known gain--speed trade-off. Here, we demonstrate that engineering the internal morphology of PEDOT:PSS:PEI films through pH-induced aqueous phase separation provides an effective route to mitigate this limitation. The resulting interconnected pore network promotes electrolyte penetration and increases the electrochemically addressable volume, while DMSO treatment and annealing enhance the continuity and ordering of the PEDOT-rich electronic phase. Consequently, porous OECT channels achieve a transconductance of 30~mS and a response time of 13~ms at an ultralow gate voltage of 0.05~V despite channel thicknesses exceeding 100~µm. Comparison of films with comparable electronic conductivity but different pore architectures identifies morphology as the dominant factor governing device performance, supporting a transition from predominantly surface-limited modulation toward spatially distributed mixed ionic--electronic transport. Beyond demonstrating a scalable water-based fabrication strategy, this work establishes internal morphology as a design parameter that complements molecular structure and device geometry in organic mixed conductors, providing a general framework for the development of high-performance OECTs, soft bioelectronics, and future neuromorphic materials.
\end{abstract}


\section{Introduction}

Organic electrochemical transistors (OECTs) convert ionic perturbations in an electrolyte into changes in the electronic conductivity of an organic semiconductor, enabling the amplification of weak biological and chemical signals into readily measurable electrical outputs \cite{Yaghmazadeh2011}. This direct coupling between ionic and electronic transport has established OECTs as versatile platforms for electrochromic displays \cite{AnderssonErsman2019}, wearable bioelectronics \cite{Marquez2020}, and chemical and biological sensing \cite{Yao2023}. Since the first OECT based on an electropolymerized polypyrrole film was introduced by White et al. in 1984 \cite{Kittlesen1984}, substantial progress in conducting-polymer synthesis and processing has improved device performance and manufacturability. In particular, aqueous dispersions of poly(3,4-ethylenedioxythiophene):polystyrene sulfonate (PEDOT:PSS) have become widely used because PEDOT:PSS combines high electronic conductivity, electrochemical stability in its oxidized state, and compatibility with solution-based fabrication \cite{Gangopadhyay2014}; \cite{Hosseini2020}.

A typical OECT consists of source (S), drain (D), and gate (G) electrodes. A conducting-polymer channel connects the source and drain, allowing a drain current (I\textsubscript{D}) to flow under an applied drain voltage (V\textsubscript{D}). The gate electrode is coupled to the channel through an electrolyte. Applying a gate voltage (V\textsubscript{G}) drives ions into or out of the channel and modulates the oxidation state and electronic conductivity of PEDOT:PSS through electrochemical doping and dedoping. The resulting change in I\textsubscript{D} constitutes the transistor output. The efficiency of this ion-to-electron conversion is commonly described by the transconductance (g\textsubscript{m}), defined as the variation in I\textsubscript{D} with V\textsubscript{G}. A high g\textsubscript{m} therefore indicates that a small ionic or electrochemical perturbation produces a large electronic current response \cite{Rivnay2013}; \cite{Ghittorelli2018}.

OECT performance is governed not only by the molecular and electronic properties of the channel material, but also by its geometry and internal morphology. At the device level, $g_m$ increases with channel width ($W$) and thickness ($d$) and decreases with channel length ($L$) \cite{Rivnay2015}; \cite{AitYazza2021}. The corresponding channel conductance is expressed as

\begin{equation}
    G = \sigma \cdot \frac{W d}{L}
\end{equation}

where $\sigma$ denotes the intrinsic electronic conductivity of the channel material. Because $g_m$ reflects the gate-induced change in conductance, $\Delta G/\Delta V_G$, increasing the electrically and electrochemically active channel volume can enhance amplification. However, increasing channel thickness also increases the distance over which ions must migrate. In dense films, this progressively restricts electrochemical modulation to regions near the electrolyte--polymer interface and slows the device response, producing the well-established trade-off between transconductance and switching speed \cite{Friedlein2018}. Conventional fabrication methods, including spin coating and inkjet printing \cite{Lin2012}; \cite{AlChamaa2022}, are well suited to planar thin-film OECTs but provide limited control over the internal transport architecture of thicker channels.

This limitation has shifted attention from external device geometry alone toward the internal morphology of organic mixed ionic--electronic conductors. Porous architectures can reduce the effective ionic diffusion length by allowing the electrolyte to penetrate the channel through interconnected pathways. At the same time, their internal surface area increases the fraction of the conducting polymer that is accessible for electrochemical doping and dedoping. Porosity therefore introduces an additional materials-design parameter: it can enhance ionic accessibility without requiring the entire channel thickness to be traversed by solid-state diffusion. Early examples include ice-templated three-dimensional macroporous PEDOT scaffolds \cite{Wan2015} and nanofiber-based OECT channels \cite{Lee2021}. Subsequent strategies have broadened this morphological toolkit, including breath-figure porous semiconducting polymers \cite{Huang2021}, nanoporous conjugated-polymer aerogel films \cite{Hu2024}, and freeze-dried porous PEDOT:PSS channels \cite{Ito2025}. These studies demonstrate that internal porosity can accelerate ion doping and raise absolute transconductance, although fabrication routes based on freeze-drying, aerogel processing, or specialized pore-forming protocols can complicate scalable aqueous processing and reproducible device integration. Critically, porosity is not an intrinsically amplifying design feature. Using chemically intact, regularly templated porous channels across conjugated polymers of different backbone ordering and side-chain polarity, Yang et al.\ showed that porous morphologies improve OECT performance only in composition-dependent regimes, with the benefit governed by the interplay of mobility, volumetric capacitance, and the effective doped volume rather than by pore presence alone \cite{Yang2026}. Complementary phase-separation studies reach a related conclusion: introducing pores can increase volumetric capacitance while leaving the mobility--capacitance product largely unchanged if electronic continuity and ion-injection pathways are not co-optimized \cite{Tang2026}.

Recent work has extended this concept to three-dimensional hydrogel semiconductors. Liu et al. showed that a porous secondary hydrogel network can guide the formation of a continuous PEDOT:PSS electronic phase while simultaneously facilitating ionic transport, thereby preserving volumetric electrochemical modulation at thicknesses extending to the millimeter scale \cite{Liu2025}. Their results demonstrate that neither maximum electronic conductivity nor maximum porosity alone is sufficient. Instead, effective three-dimensional modulation requires a balance between continuous electronic percolation and electrolyte-accessible ionic pathways. This establishes internal morphology as a central determinant of ion--electron coupling in thick mixed conductors. However, scalable routes for generating such transport architectures outside permanently cross-linked hydrogel networks remain comparatively underexplored, particularly for freestanding polyelectrolyte-complex films and conventional OECT configurations.

Aqueous phase separation (APS), originating from membrane science, provides a distinct route for engineering such internal morphology. APS enables porous polyelectrolyte films to be formed through the controlled complexation and precipitation of oppositely charged polymers \cite{Baig2021}; \cite{Sadman2019}; \cite{HaqueMizan2023}; \cite{Restrepo2024}. The driving force for polyelectrolyte-complex formation is largely associated with the entropy gained through the release of counterions \cite{michaels1965polyelectrolyte}. During solution preparation, premature precipitation is suppressed by weakening electrostatic interactions between the polymers \cite{Wang2014}. For strong polyelectrolytes, which remain charged over a broad pH range, this can be achieved by increasing ionic strength through salt addition \cite{Sadman2019}; \cite{Restrepo2024}. For weak polyelectrolytes, whose degree of ionization depends on pH, complexation can instead be delayed by adjusting the solution pH to reduce polymer charge \cite{Baig2021}; \cite{HaqueMizan2023}. Phase separation is subsequently triggered by removing salt or shifting the pH in a coagulation bath. Because polymer composition and coagulation conditions determine the kinetics of complexation and precipitation, APS provides direct control over pore formation, connectivity, and structural stability.

\begin{figure}
    \centering
    \includegraphics[width=\linewidth]{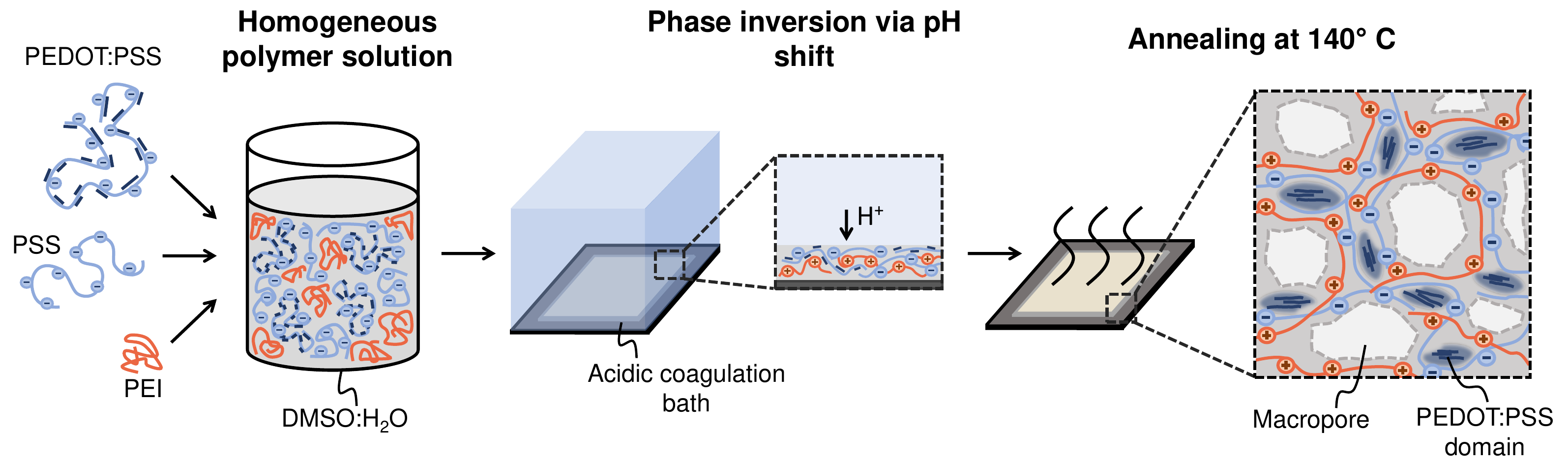}
    \caption{Schematic of the aqueous phase separation process: a homogeneous solution of PEDOT:PSS, PSS, and PEI is cast and immersed in an acidic coagulation bath. The pH shift triggers complexation and precipitation, forming a porous PEDOT:PSS:PEI film.}
    \label{fig:F1}
\end{figure}

Here, APS is applied as a morphology-engineering platform for PEDOT:PSS-based mixed conductors. PEDOT:PSS is combined with additional PSS and branched polyethyleneimine (PEI), while dimethyl sulfoxide (DMSO) is used as a co-solvent and secondary dopant. The homogeneous precursor solution is cast and immersed in an acidic coagulation bath. Protonation of the PEI amine groups induces complexation with the sulfonate groups of PSS, triggering phase separation and forming a porous PEDOT:PSS:PEI matrix (\textbf{Figure \ref{fig:F1}}). In contrast to templated hydrogel approaches, the resulting architecture is generated through pH-triggered polyelectrolyte complexation and can be produced as a freestanding, processable film.

We hypothesize that the internal pore architecture generated by pH-induced APS increases electrolyte accessibility and reduces the effective ionic transport distance throughout the channel, thereby promoting volumetric electrochemical modulation while maintaining an electronically percolated PEDOT phase. Consequently, controlled morphology should enable high transconductance and short response times to be achieved concurrently, mitigating the conventional gain--speed trade-off of thick OECT channels. To test this hypothesis, we systematically vary the PSS:PEI composition, coagulation conditions, DMSO content, and annealing treatment, and relate the resulting morphology, mechanical properties, and electronic conductivity to OECT performance. The study therefore positions APS not only as a fabrication method for porous channels, but as a platform for identifying how morphology governs mixed ionic--electronic transport in thick organic electrochemical devices.

\subsection{Morphology: Effect of PSS-PEI ratio and acidic bath}\label{sec:morphology}

\begin{figure}[h!]
    \centering
    \includegraphics[width=\textwidth]{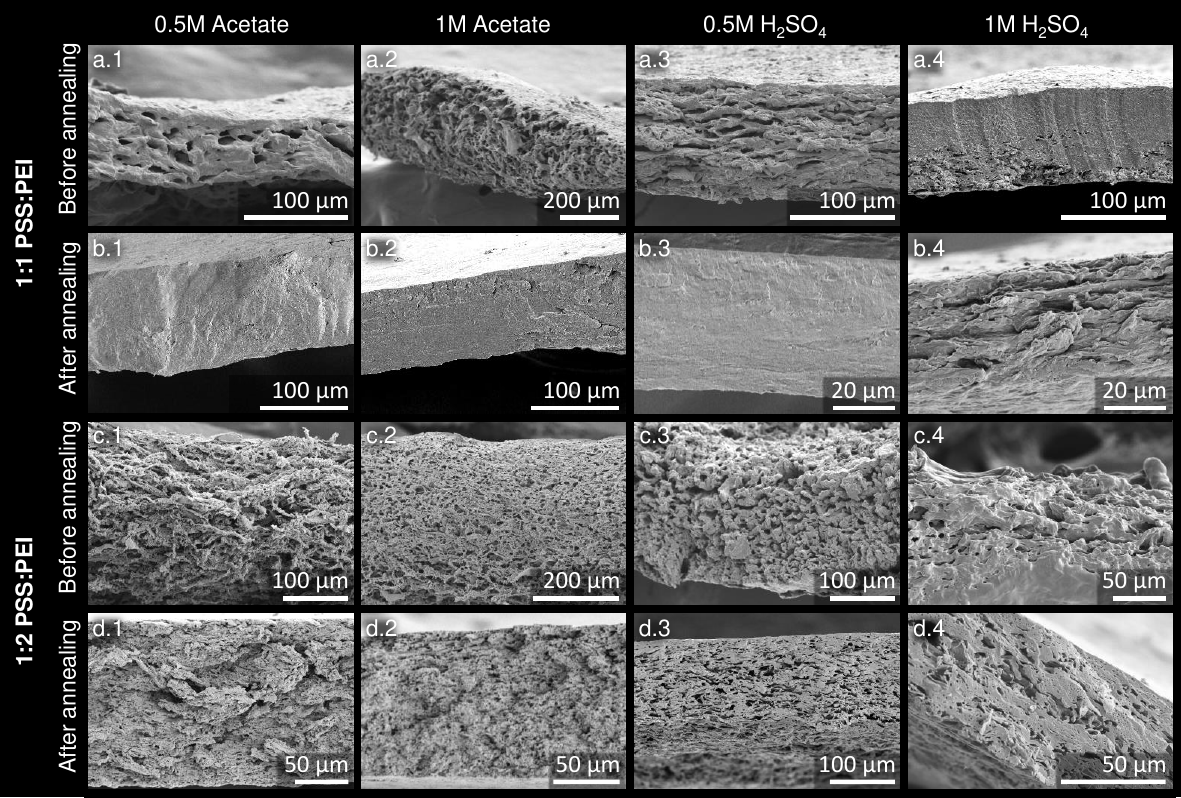}
    \caption{Cross-sectional SEM images of PEDOT:PSS:PEI films. Films with PSS:PEI ratios of 1:1 and 1:2 were coagulated in 0.5 M or 1 M acetate buffer or H$_2$SO$_4$ baths before and after annealing at \SI{140}{\celsius}. All samples were immersed in water and freeze-dried after fabrication to preserve the hydrated pore structure during imaging.}
    \label{fig:SEM-CS}
\end{figure}

The internal morphology of polyelectrolyte-complex films is governed by the kinetics and extent of complexation during coagulation. Previous work has shown that the composition and acidity of the coagulation bath strongly influence pore formation in PEI:PSS systems \cite{Baig2021}. To examine how these parameters control the transport architecture of PEDOT:PSS:PEI films, compositions with PSS:PEI ratios of 1:1 and 1:2 were cast and coagulated in 0.5 and \SI{1}{M} acetate buffer or H$_2$SO$_4$ baths. The films were subsequently annealed at \SI{140}{\celsius} for \SI{90}{\minute}. The resulting cross-sectional morphologies are shown in \textbf{Figure~\ref{fig:SEM-CS}}.

SEM analysis reveals substantial differences in pore morphology as a function of both polymer composition and coagulation conditions. Films prepared from 1:1 and 1:2 precursor solutions exhibit nonuniform thicknesses ranging from approximately 80 to 200~µm. Samples coagulated in 0.5 and 1~M acetate buffer or in \SI{0.5}{M} H$_2$SO$_4$ display comparatively large pores, whereas coagulation in \SI{1}{M} H$_2$SO$_4$ produces a more compact and homogeneous structure. After annealing and rehydration, the 1:1 films largely lose their distinct porous morphology. In contrast, the 1:2 films retain a clearly resolved pore network and exhibit improved structural homogeneity after annealing.

Importantly, the cross-sectional images reveal openings between neighboring pores (Fig.~2d, inset), indicating that the pore space is interconnected rather than composed of isolated voids. This distinction is central to the transport function of the films. Interconnected pores can form continuous electrolyte-accessible pathways through the channel, thereby reducing the effective distance over which ions must diffuse through the polymer-rich phase. The relevant structural parameter is therefore not pore size alone, but the connectivity and continuity of the pore network. This interpretation is consistent with recent work on three-dimensional hydrogel semiconductors, where efficient electrochemical modulation was shown to require simultaneous continuity of ionic and electronic transport pathways rather than maximum porosity alone \cite{Liu2025}.

The morphological differences between the 1:1 and 1:2 films can be rationalized by their polyelectrolyte complexation behavior. An excess of PEI has previously been shown to promote precipitation and stabilization of PSS-containing complexes \cite{Baig2021}; \cite{HaqueMizan2023}. Stable polyelectrolyte complexes are generally favored when the positive and negative charges approach stoichiometric compensation \cite{Chen2021}. In an idealized description, each negatively charged sulfonate group of PSS would be compensated by one protonated amine group of PEI. However, branched PEI contains primary, secondary, and tertiary amine groups with pKa values of 4.5, 6.7, and 11.6, respectively \cite{Willner1993}. The nominal PSS:PEI ratios therefore overestimate the number of amine groups that effectively participate in complexation.

Studies of DNA/PEI complexes suggest that primary and secondary amines contribute most strongly to complex formation \cite{Choosakoonkriang2003}; \cite{Hu2014}, because they are more accessible to the surrounding solvent and more readily protonated \cite{Harden1986}. The PEI used here contains 31\% primary, 39\% secondary, and 30\% tertiary amine groups \cite{VonHarpe2000}. If only primary and secondary amines are considered effective complexation sites, the nominal PSS:PEI ratios of 1:1 and 1:2 correspond to effective charge ratios of approximately 1:0.7 and 1:1.4, respectively. The 1:1 composition therefore retains uncompensated negative charge, whereas the 1:2 composition provides a slight excess of accessible amine groups and approaches more complete charge compensation.

This interpretation is consistent with the macroscopic behavior of the films. The 1:1 samples are soft and undergo pronounced swelling in water, indicating incomplete charge compensation and a higher density of osmotically active fixed charges \cite{Chen2021}. Their loss of pore structure and reduction in thickness after annealing may result from enhanced chain mobility, structural collapse, and material loss associated with incomplete complexation \cite{Fu2017}. By contrast, the more highly compensated 1:2 films retain their porous architecture after annealing and rehydration.

The effect of the coagulation bath further supports this charge-compensation mechanism. In acetate buffer at pH~4, protonation of PEI is incomplete, particularly for the less basic amine populations. In both 0.5~M and 1~M H$_2$SO$_4$ solutions (pH~$<$~1), PEI is protonated more extensively, increasing its interaction with PSS and stabilizing the polyelectrolyte network. The \SI{1}{M} H$_2$SO$_4$ condition therefore produces the most structurally integrated films. These results show that polymer stoichiometry and protonation conditions jointly determine whether APS generates a mechanically unstable porous morphology or an interconnected structure capable of supporting coupled ionic and electronic transport.

\subsection{Electrical Conductivity}

Efficient operation of mixed ionic--electronic conductors requires a continuous electronic transport network in parallel with electrolyte-accessible ionic pathways. Porosity can improve ion transport, but excessive disruption of the PEDOT-rich phase may reduce electronic percolation. The relevant design objective is therefore not to maximize porosity or conductivity independently, but to establish a morphology in which both charge-carrier pathways remain continuous. This balance has recently been identified as a central requirement for three-dimensional hydrogel semiconductors, where increasing porosity enhances ionic conductivity but eventually compromises electronic conductivity and electrochemical modulation \cite{Liu2025}.

Among conducting polymers, PEDOT combines relatively high electrochemical stability with an electrical conductivity that can be adjusted over a broad range (10\textsuperscript{-3}--10\textsuperscript{3} S cm\textsuperscript{-1}) compared with PPy and PANI. In PEDOT:PSS:PEI films, the electronic conductivity determines whether the PEDOT-rich domains form a sufficiently continuous pathway for charge transport through the channel. At the same time, this pathway must remain accessible to ions during electrochemical doping and dedoping.

DMSO is widely used to modify the phase structure and conductivity of PEDOT:PSS. Previous studies suggest that DMSO weakens interactions between PEDOT and the insulating PSS-rich shell, inducing a transition from coiled to more extended PEDOT conformations \cite{Yildirim2018}. This promotes chain disentanglement, ordering, and interchain charge transport \cite{Ouyang2004}. However, excessive DMSO can produce enlarged PEDOT-rich domains and brittle intergranular interfaces, ultimately reducing mechanical integrity and conductivity \cite{Lee2016}. Electrical conductivity measurements ($n = 3$ for each condition) were therefore performed to determine the effects of coagulation bath, annealing treatment, and PSS:PEI ratio on the electronic transport properties of the composite films. Preliminary experiments identified 25~wt.\% DMSO as the optimum concentration for the present formulation (\textbf{Supporting Information S1}).

\begin{figure}[h!]
    \centering
    \includegraphics[width=0.5\textwidth]{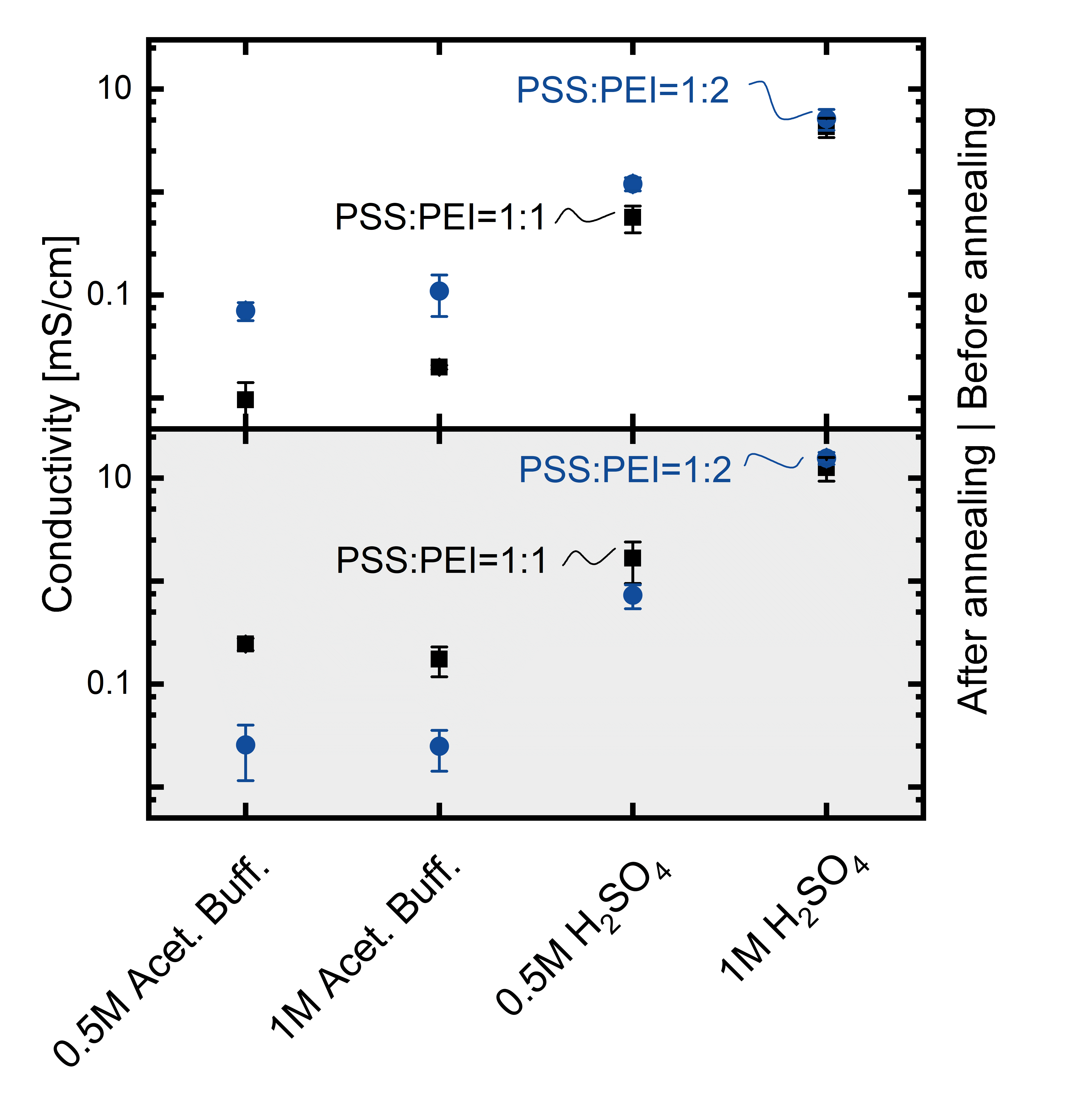}
    \caption{Electrical conductivity of PEDOT:PSS:PEI films as a function of coagulation bath and PSS:PEI ratio. All films were prepared with 25 wt.\% DMSO. Each point represents the mean of three independent measurements, and the error bars denote the corresponding standard deviation.}
    \label{fig:Conductivity}
\end{figure}

\textbf{Figure \ref{fig:Conductivity}} shows the conductivity of the PEDOT:PSS:PEI films as a function of PSS:PEI ratio, coagulation bath, and annealing treatment. The measured values range from approximately \SI{0.01}{mS/cm} to \SI{15}{mS/cm}. These conductivities are lower than those typically reported for pure PEDOT:PSS films \cite{Jiang2020}, which is expected because the composite contains substantial fractions of electronically insulating PSS and PEI. Nevertheless, the measured values demonstrate that the PEDOT-rich phase remains electronically percolated across the macroscopic film.

Electronic transport in conducting-polymer composites occurs through a combination of transport along conjugated domains and tunneling or hopping between neighboring conducting regions \cite{Mondal2021}. In PEDOT, interchain charge transfer frequently limits the overall carrier mobility \cite{Kim2019}. The reduced conductivity of the present films can therefore be attributed to the interruption or separation of PEDOT:PSS domains by insulating PEI- and PSS-rich regions. Within experimental uncertainty, the PSS:PEI ratio does not produce a significant change in conductivity. This observation is important for interpreting the device results: the pronounced differences between the 1:1 and 1:2 films cannot be explained by electronic conductivity alone and are therefore more directly associated with differences in internal morphology and ionic accessibility.

The highest conductivities are obtained for films coagulated in \SI{1}{M} H$_2$SO$_4$. The stronger acidic environment promotes more complete protonation of PEI and more extensive complexation with PSS, including PSS associated with the PEDOT:PSS dispersion. In addition, treatment with H$_2$SO$_4$ may remove part of the excess PSS-rich phase and promote closer packing and crystallization of PEDOT chains through enhanced $\pi$--$\pi$ interactions \cite{Gueye2020}. This condition therefore appears to improve both structural consolidation of the polyelectrolyte matrix and electronic continuity of the PEDOT-rich phase.

Annealing increases the conductivity by approximately a factor of three, from \SI{5}{mS/cm} to \SI{15}{mS/cm}. Removal of residual water and processing solvents reduces separation between conducting domains and promotes reorganization of the polymer matrix. Annealing can additionally enhance PEDOT crystallinity and molecular ordering \cite{Carter2023}. The conductivity increase is therefore attributed to improved interchain coupling and continuity of the PEDOT-rich network rather than to a change in the overall polymer composition.

DMSO also modifies the mechanical properties of the PEDOT:PSS:PEI films. The corresponding tensile-test data are presented in \textbf{Supporting Information S2} and \textbf{Supporting Information S3}. Incorporation of 25~wt.\% DMSO produces more pliable films than the corresponding 0~wt.\% formulations. Mechanical characterization ($n = 3$ for each condition) shows an increased strain at break in the hydrated state, particularly for DMSO-containing samples. This behavior is consistent with the plasticizing effect of water in hydrated polyelectrolyte complexes, where water weakens intermolecular interactions and increases chain mobility \cite{Schaaf2015}.

DMSO therefore performs two complementary functions. First, it acts as a processing aid that redistributes or partially removes the insulating PSS-rich phase, improving electronic connectivity \cite{Yildirim2018}. Second, it promotes more extended PEDOT conformations and improved intermolecular ordering \cite{Ouyang2004}. The resulting films combine an electronically percolated PEDOT network with enhanced mechanical compliance under hydrated conditions. This combination is particularly relevant for OECT operation, because the channel must sustain electronic transport while accommodating swelling and ion penetration during repeated electrochemical modulation.

\subsection{OECTs Characterization}

\subsubsection{Operating principles}

\begin{figure}[h!]
    \centering
    \includegraphics[width=0.9\textwidth]{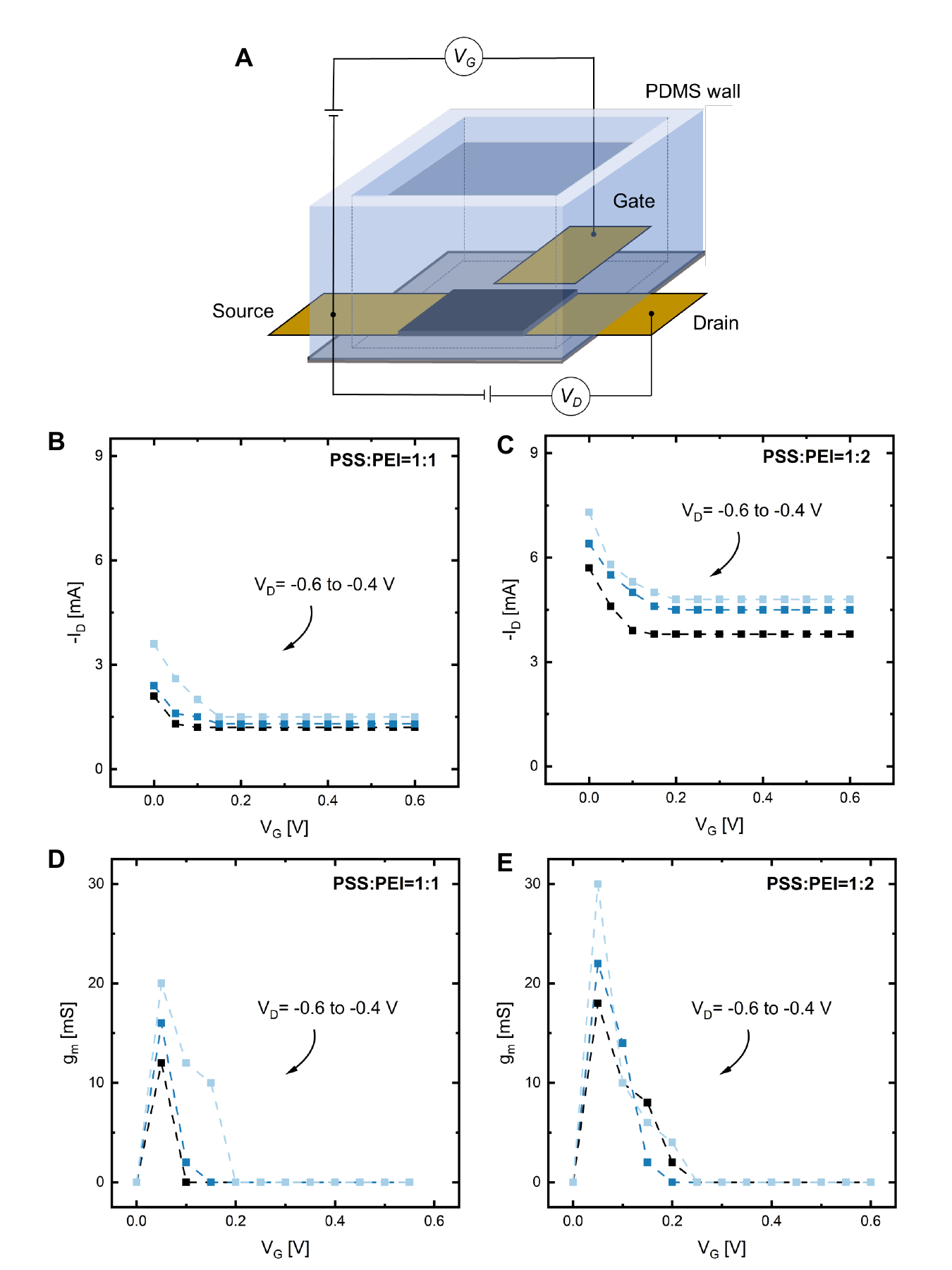}
    \caption{A) Schematic of the OECT measurement cell. B--C) Transfer characteristics and D--E) corresponding transconductance (g\textsubscript{m}) of PEDOT:PSS:PEI channels with different PSS:PEI ratios.}
    \label{fig:OECT}
\end{figure}

OECTs employ a three-terminal architecture comprising source, drain, and gate electrodes. The source and drain are connected through the mixed-conducting channel, whereas the gate is coupled to the channel through an electrolyte, as illustrated in \textbf{Figure \ref{fig:OECT} A}.

Application of a positive or negative gate voltage (V\textsubscript{G}) drives ions from the electrolyte into the channel and changes the oxidation state of the conducting polymer. This electrochemical doping or dedoping modifies the channel conductivity, such that relatively small variations in V\textsubscript{G} can produce pronounced changes in the drain current (I\textsubscript{D}). The corresponding transconductance (g\textsubscript{m}) quantifies the efficiency of this gate-controlled current modulation \cite{Tian2022}.

For PEDOT:PSS channels, application of a positive gate voltage drives cations into the polymer and reduces the oxidized, highly conducting PEDOT$^+$ state to the less conducting neutral PEDOT$^0$ state. This reversible dedoping process can be represented as

\begin{equation}
    \mathrm{PEDOT^+:PSS^- + M^+ + e^- \rightleftharpoons PEDOT^0 + M^+:PSS^-}
\end{equation}

where $M^+$ denotes a mobile cation in the electrolyte and $e^-$ an electron supplied through the source electrode. Because the negatively charged PSS chains remain immobilized within the polymer matrix, charge compensation occurs through cation uptake. Upon removal or reversal of the gate potential, the cations migrate back toward the electrolyte and the higher-conductivity state of PEDOT:PSS is restored \cite{Nikolou2008}.

\subsubsection{Transfer Curve}

The transfer curve describes the drain current (I\textsubscript{D}) as a function of gate voltage (V\textsubscript{G}) at a fixed drain voltage (V\textsubscript{D}). It therefore provides a direct measure of how efficiently electrochemical modulation of the channel is converted into an electronic output.

\textbf{Figure \ref{fig:OECT} B--C} show the transfer characteristics of the PSS:PEI = 1:1 and 1:2 devices for drain voltages ranging from $-0.6$ to $-0.4$~V, while V\textsubscript{G} was swept from 0 to 0.6~V in 0.05~V increments. In both devices, I\textsubscript{D} decreases with increasing V\textsubscript{G}, confirming depletion-mode operation. The decrease results from the ingress of Na\textsuperscript{+} ions into the channel, which compensates the PSS charge, reduces the hole density in PEDOT, and lowers its electronic conductivity \cite{AitYazza2021}.

The PSS:PEI = 1:2 device reaches a maximum drain current of 7.3~mA, approximately twice that of the 1:1 device (3.6~mA). Because both devices have identical macroscopic geometries, this difference indicates a higher effective channel conductance for the 1:2 film. The conductivity measurements alone do not fully account for this increase, because the two compositions exhibit comparable electronic conductivities within experimental error. The higher drain current is therefore consistent with a more favorable internal morphology that preserves electronic continuity while improving electrolyte access to the channel volume.

\subsubsection{Transconductance g\textsubscript{m}}

The transconductance describes the sensitivity of the drain current to the applied gate voltage and is defined as the first derivative of the transfer characteristic:

\begin{equation}
    g\textsubscript{m} = \frac{\partial I_{D}}{\partial V_{G}}
\end{equation}

In PEDOT:PSS-based OECTs, g\textsubscript{m} reflects the coupling between ionic charge injection and the resulting modulation of electronic transport. Ions entering the channel alter the electrostatic environment of the PSS sulfonate groups and thereby change the concentration and mobility of holes in the PEDOT-rich phase. A high g\textsubscript{m} therefore requires both efficient ionic access to the active material and a continuous electronic pathway through which the resulting conductance change can be measured.

\textbf{Figure \ref{fig:OECT} D--E} present the transconductance derived from the transfer curves for three drain voltages. The PSS:PEI = 1:2 film reaches a maximum transconductance of $30 \pm 6.4\ \mathrm{mS}$ at $V_\mathrm{G} = 0.05\ \mathrm{V}$ and $V_\mathrm{D} = -0.6\ \mathrm{V}$ ($N = 5$). Under the same conditions, the PSS:PEI = 1:1 film reaches $18 \pm 5.3\ \mathrm{mS}$ ($N = 5$), corresponding to an increase of approximately 66.7\% for the 1:2 composition.

The difference in transconductance cannot be attributed solely to electronic conductivity, because the conductivity values of the two compositions overlap within experimental uncertainty. Instead, the higher g\textsubscript{m} of the 1:2 device is consistent with its more stable and interconnected pore network. This morphology increases the fraction of PEDOT:PSS that is accessible to electrolyte ions and thereby increases the conductance change $\Delta G$ generated per unit gate-voltage variation $\Delta V\textsubscript{G}$.

The interconnected pore structure is expected to contribute in two complementary ways. First, the hydrated pore network provides continuous pathways for Na$^+$ transport through the film, reducing the effective solid-state diffusion distance. Second, the internal pore surface exposes a larger fraction of PEDOT-rich domains to the electrolyte, increasing the electrochemically addressable volume. Together, these effects promote spatially distributed doping and dedoping throughout the channel rather than restricting charge modulation to the external film surface.

This interpretation is consistent with the broader principle established for three-dimensional mixed conductors: high electrochemical performance requires simultaneous continuity of ionic and electronic transport pathways. Increasing porosity alone can improve ionic mobility but may disrupt electronic percolation, whereas a dense electronic network may impede electrolyte penetration. Material-agnostic porous-channel studies reinforce the same point by showing that porosity benefits are chemistry-dependent and must be evaluated through mobility, volumetric capacitance, and the electrochemically doped volume rather than pore geometry in isolation \cite{Yang2026}. The superior performance of the 1:2 film therefore suggests that its morphology provides a more favorable balance between these two transport modes.

The incorporation of PEI may further support ion transport by increasing water uptake and lowering the elastic modulus of the hydrated matrix. The resulting swelling facilitates penetration of ions and solvent molecules during electrochemical modulation and can improve ion-to-electron signal conversion \cite{Moser2019}. DMSO treatment and annealing, in turn, improve PEDOT ordering and electronic connectivity. The enhanced device response should therefore be understood as the combined result of ionic accessibility, electronic continuity, and mechanical accommodation of repeated swelling and deswelling.

The high transconductance observed at the low gate voltage of V\textsubscript{G} = 0.05~V is particularly relevant for bioelectronic applications, where excessive electrical stimulation must be avoided. This operating regime is consistent with recent reports of low-voltage OECT operation \cite{Yan2024} and demonstrates that the porous channels can generate substantial electronic amplification under mild electrochemical driving.

The present results support a transition from predominantly surface-limited modulation toward more spatially distributed mixed ionic--electronic conduction. However, the extent of volumetric modulation cannot be established from transfer characteristics alone. Systematic studies of regularly patterned porous channels and of phase-separated versus selectively dissolved OMIECs show that pore formation and amplification need not increase in parallel and that the benefit of porosity depends on polymer chemistry and on the resulting doped volume \cite{Yang2026}; \cite{Tang2026}. These findings reinforce the need for quantitative transport metrics beyond transfer curves. Direct determination of ionic conductivity, electronic conductivity, volumetric capacitance, and diffusion impedance by electrochemical impedance spectroscopy or thickness-dependent capacitance measurements would provide quantitative validation of the proposed transport mechanism.

\subsubsection{Trade-off}

OECT switching dynamics are governed by the coupled transport of ions and electronic charge carriers. Increasing the channel thickness increases the quantity of electrochemically active material and can therefore enhance g\textsubscript{m}. In dense channels, however, it also increases the ion-transport distance and slows the response. This gives rise to the characteristic trade-off between amplification and switching speed. Conventional OECTs address this limitation by employing channels thinner than 1~µm. Optimized devices with thicknesses from several tens to several hundred nanometers typically exhibit g\textsubscript{m} values of approximately 5~mS and response times in the tens-of-milliseconds range \cite{AitYazza2021}; \cite{Khodagholy2013}.

\begin{figure}[h!]
    \centering
    \includegraphics[width=1\textwidth]{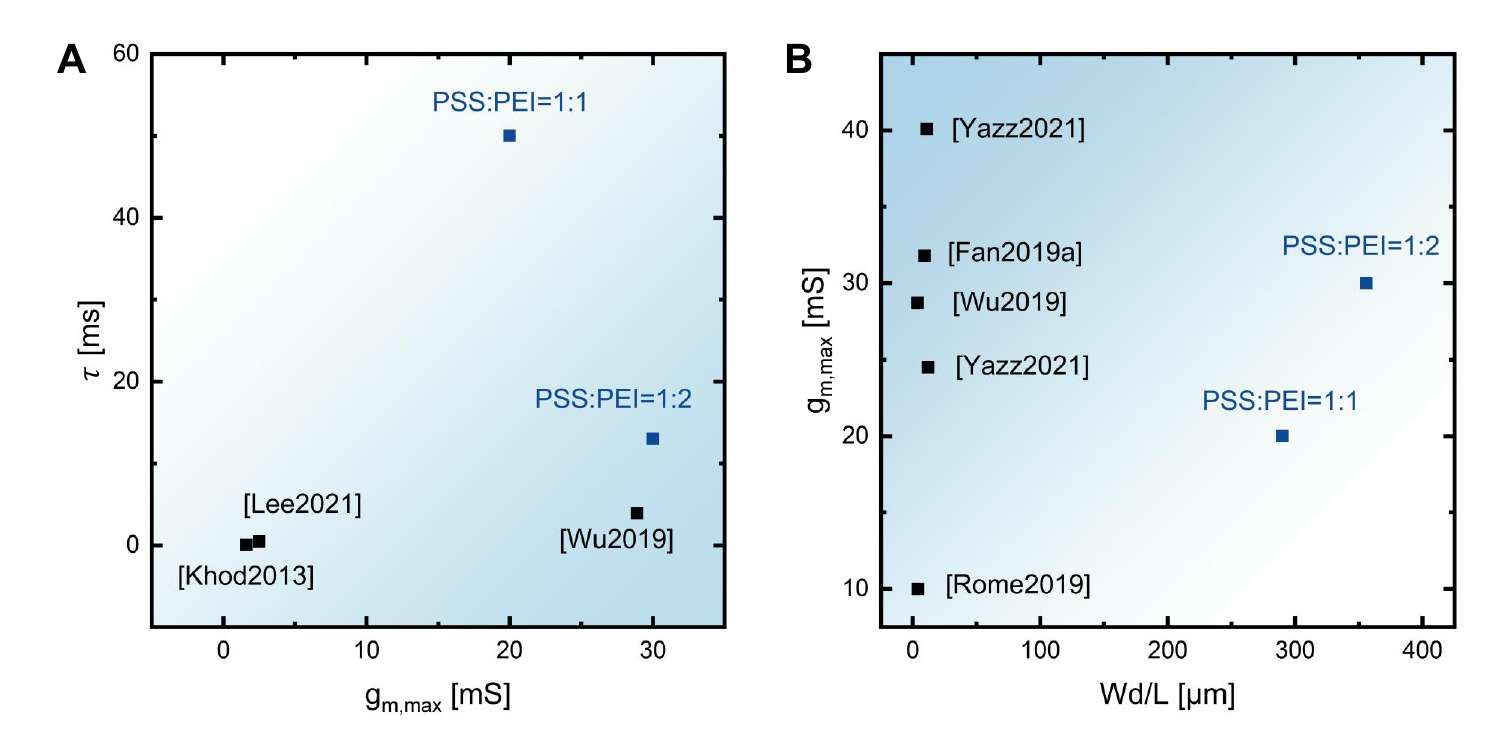}
    \caption{A) Trade-off between response time and maximum transconductance. Literature values are shown in black \cite{Lee2021}; \cite{Khodagholy2013}; \cite{Wu2019}. B) Comparison of maximum transconductance normalized by channel geometry with literature values adapted from \cite{AitYazza2021}; \cite{Wu2019}; \cite{Romele2019}; \cite{Fan20193d}.}
    \label{fig:trade-off}
\end{figure}

To evaluate this trade-off, \textbf{Figure \ref{fig:trade-off} A} compares the response time ($\tau$) at maximum transconductance (g\textsubscript{m,max}) for the PSS:PEI = 1:1 and 1:2 devices with values reported in the literature. The comparison shows that both porous films achieve performance within the range of previously reported devices, despite their substantially greater thickness. The PSS:PEI = 1:2 device provides the more favorable combination of high g\textsubscript{m,max} and short $\tau$, consistent with more efficient ion penetration into the channel \cite{Sheliakina2018}.

The dry PSS:PEI = 1:1 and 1:2 films have thicknesses of approximately 120--200~µm and are therefore substantially thicker than conventional OECT channels. Nevertheless, they exhibit response times of 39~ms and 13~ms, respectively. These values demonstrate that the response is not governed by the full macroscopic film thickness alone. Instead, the interconnected pore network appears to divide the channel into shorter local ion-transport distances, allowing electrolyte access throughout the structure and promoting more homogeneous electrochemical modulation \cite{Guo2023}.

The faster response of the 1:2 device likely arises from the combined effects of its electronically continuous PEDOT-rich phase and its more persistent pore network. Electronic conductivity supports rapid redistribution of holes through the channel, whereas the interconnected pore phase supports rapid cation ingress and egress. The 1:1 film, by contrast, partially loses its porous structure after annealing and rehydration, increasing the effective ionic transport distance and slowing equilibration.

This morphology therefore mitigates, rather than fully eliminates, the conventional gain--speed trade-off. The porous channel retains a large electrochemically active volume while reducing the local ion-diffusion length. In this sense, it combines the charge-storage capacity associated with a thick channel with response dynamics more commonly associated with thinner films \cite{Li2024}. A definitive demonstration of complete volumetric modulation would nevertheless require thickness-dependent capacitance or impedance measurements.

The influence of device geometry is further assessed through the normalized transconductance, g\textsubscript{m,n} = g\textsubscript{m,max}/(W d L\textsuperscript{-1}), shown in \textbf{Figure \ref{fig:trade-off} B}. Whereas literature values frequently exceed 2~mS/µm, the present devices exhibit lower values of 0.069--0.08~mS/µm. This difference reflects the non-optimized, macroscopic device geometry and the large channel thickness rather than an intrinsic limitation of the material.

The normalized values should therefore be interpreted separately from the principal structure--function comparison of this study. The high absolute transconductance demonstrates that a large fraction of the porous channel contributes to charge modulation. Further optimization of $W$, $L$, and $d$ could improve the geometry-normalized performance without changing the underlying material concept.

The most direct evidence for the role of morphology is provided by the comparison between the PSS:PEI = 1:1 and 1:2 films coagulated in 1~M H$_2$SO$_4$. These films exhibit comparable electronic conductivity (\textbf{Figure~\ref{fig:Conductivity}}) but clearly different pore structures (\textbf{Figure~\ref{fig:SEM-CS}}). The 1:2 film retains a more interconnected porous network after annealing and rehydration and simultaneously exhibits higher transconductance and a shorter response time. This internal comparison minimizes differences in device geometry and processing and therefore identifies morphology as the dominant variable underlying the performance enhancement.

Taken together, the results support the central hypothesis that APS-generated pore connectivity improves ionic access to the PEDOT-rich phase while preserving electronic percolation. The resulting morphology increases the electrochemically addressable channel volume and shortens effective ionic transport distances, thereby producing a larger conductance modulation and faster response. The current measurements establish the functional relevance of this morphology, while future impedance, capacitance, and ion-transport measurements should quantify the degree of volumetric mixed ionic--electronic conduction.

\section{Conclusions}

This work establishes pH-induced aqueous phase separation as a versatile route for engineering the internal morphology of PEDOT:PSS:PEI mixed conductors. By varying the PSS:PEI ratio and coagulation conditions, the pore structure and structural stability of the films could be systematically adjusted, while DMSO incorporation and annealing independently modified the electronic transport properties of the PEDOT-rich phase. The resulting PEDOT:PSS:PEI films reached conductivities of up to 15.65~mS/cm after annealing and retained mechanical compliance in the hydrated state, demonstrating that morphology, electronic conductivity, and mechanical properties can be tuned through complementary processing parameters.

OECT characterization reveals that the PSS:PEI = 1:2 composition provides the most favorable balance between amplification and response dynamics, achieving a transconductance of 30~mS and a response time of 13~ms. These values are comparable to those of substantially thinner OECT channels despite the macroscopic film thickness of 120--200~µm. The performance enhancement is attributed to the combined effects of improved electronic continuity, promoted by DMSO treatment and annealing, and a persistent interconnected pore network that facilitates ion penetration into the channel.

The comparison between the PSS:PEI = 1:1 and 1:2 films provides the clearest evidence for the role of internal morphology. Although both compositions exhibit comparable electronic conductivity under the selected coagulation conditions, the 1:2 film retains a more interconnected porous architecture after annealing and rehydration and simultaneously exhibits higher transconductance and faster switching. These results support the interpretation that pore connectivity increases the electrochemically accessible fraction of the PEDOT-rich phase and reduces the effective ion-transport distance within the channel.

Accordingly, the present findings indicate a shift from predominantly surface-limited modulation toward more spatially distributed mixed ionic--electronic transport. The porous structure enables electrolyte access throughout the film and supports electrochemical modulation over a larger channel volume than would be accessible in a dense film of comparable thickness. This morphology therefore mitigates the conventional gain--speed trade-off by combining a large electrochemically active volume with shortened local ion-diffusion pathways. However, direct confirmation of complete volumetric modulation will require quantitative measurements of volumetric capacitance, ionic conductivity, and diffusion impedance.

More broadly, the study identifies internal morphology as a design parameter that complements molecular structure and device geometry in organic mixed conductors. Consistent with recent assessments of the true role of porosity in OECTs \cite{Yang2026}, the APS process offers a scalable, water-based platform for controlling pore formation through polyelectrolyte complexation and provides a route toward systematic investigation of how pore size, connectivity, tortuosity, and polymer-phase continuity govern electrochemical device behavior. Future studies combining electrochemical impedance spectroscopy, thickness-dependent capacitance measurements, three-dimensional tomography, and in situ swelling analysis should enable extraction of the intrinsic figure of merit ($\mu C^*$) and establish quantitative structure--transport--performance relationships.

By transferring concepts from membrane science to organic electronics, this work positions aqueous phase separation as a morphology-engineering strategy for mixed ionic--electronic conductors that complements existing porous-channel approaches based on freeze-drying, aerogels, breath-figure films, and hydrogel templating \cite{Huang2021}; \cite{Hu2024}; \cite{Ito2025}; \cite{Liu2025}. The resulting design framework is relevant not only to OECTs, but also to soft bioelectronic interfaces, electrochemical sensors, and future neuromorphic or reservoir-computing systems in which distributed ionic dynamics and history-dependent material states are exploited as functional resources.

\section{Experimental Section}
\subsection{Materials}
The dry re-dispersable pellets poly(3,4-ethylenedioxythiophene) polystyrene sulfonate (PEDOT:PSS) were purchased from Sigma Aldrich. PEDOT:PSS aqueous solution (Clevios PH 1000) was purchased from Heraeus and the concentration of PEDOT:PSS was 1.3\% by weight. Poly(sodium 4-styrenesulfonate) (PSS) with high molecular weight (MW) (average MW $\sim$ 1,000,000) and branched polyethyleneimine (PEI) with average MW $\sim$ 25,000 were supplied from Sigma Aldrich. The chemicals DMSO (anhydrous, $\geq$ 99.9\%), acetic acid ($\geq$ 99.7\%), sodium acetate (anhydrous, $\geq$ 99\%),  sulfuric acid (H\textsubscript{2}SO\textsubscript{4}, 95 - 98\%), Glycerin ($\geq$ 99.5\%), sodium chloride (NaCl, ACS reagent, $\geq$ 99.0\%), (3-glycidyloxypropyl)trimethoxysilane (GOPS, $\geq$ 98\%), ethylene glycol (EG, $\geq$ 99\%) and Dodecylbenzene sulfonic acid (DBSA, mixture of isomers, $\geq$ 95\%) were purchased from Sigma- Aldrich. For the poly(dimethyl siloxane) (PDMS), Sylgard 184 Silicone by Dow Corning was used. All the materials were used without further purification. 

\subsection{Film Fabrication}

For the preparation of the polymer solutions used for film casting, PEDOT:PSS pellets were first mixed with DMSO and/or water until fully homogenized using an asymmetric centrifugal mixer (SpeedMixer, Hauschild, Germany). Then, PSS and PEI were added with a monomer ratio of 1:1 or 1:2 and mixed until a homogeneous polymer solution was obtained. The ratio was calculated using the molecular weight of the PSS and PEI repeating units, which are M$_{PSS}$=\SI{205.6}{g/mol} and M$_{PEI}$= \SI{43.04}{g/mol}, respectively. The detailed composition of the solution is given in Table~\ref{tab:Polymer_composition}.

\begin{table*}[htbp]
  \centering
  \caption{Composition of the polyelectrolyte polymer solutions for film casting}
  \vspace{0.2cm}
    \begin{tabular}{c c c c c c c c}
    \toprule
     Solution & PSS:PEI & PSS & PEI & PEDOT:PSS & DMSO & H$_2$O \\
      - & monomer ratio & [wt$\%$] & [wt$\%$] & [wt$\%$] & [wt$\%$] & [wt$\%$] \\
    \midrule
     S.1 & 1:1 & 16.6 & 3.4 & 5 & 25 &  50  \\
     S.2 & 1:1 & 16.6 & 3.4 & 5 & - &  75  \\
     S.3 & 1:2 & 14.2 & 5.8 & 5 & 25 & 50 \\
     S.4 & 1:2 & 14.2 & 5.8 & 5 & - & 75 \\
    \bottomrule
    \end{tabular}
  \label{tab:Polymer_composition}
\end{table*}

For film fabrication (shown in \textbf{Figure \ref{fig:F1}}), the polymer solution was placed on top of a glass slide and then evenly spread using a casting knife with a set height of \SI{500}{\micro m}. Subsequently, the casted film was immersed in a coagulation bath consisting of a \SI{0.5}{M} or \SI{1}{M} acetate buffer solution with a pH of 4, or \SI{0.5}{M} or \SI{1}{M} sulfuric acid solution. After \SI{10}{min} in the coagulation bath, the films were transferred to DI water for 3h to remove excess acid. Finally, overnight, the films were immersed in a \SI{50}{wt\%} glycerol/water solution and were then air-dried for further characterization. Films were annealed in an oven (Binder, Aldrich) at \SI{140}{\degreeCelsius} for \SI{90}{\minute}.

\subsection{Scanning Electron Microscopy}
The morphologies of the PEDOT:PSS:PEI films under dry and wet states were analyzed by field emission scanning electron microscopy (FeSEM, Hitachi SU5000). A freeze-drying method was employed to obtain the cross-section of the films. The samples were first freeze-dried in the liquid nitrogen and subsequently fractured to achieve a planar cross-sectional fracture. Before FeSEM imaging, all specimens were lyophilized for 2 days (Martin Christ, Alpha 1-4 Dplus) and sputter-coated with a 6 nm thin layer of Au/Pd using a sputter coater (Leica EM ACE600). An accelerating voltage of 5 kV and 8 to 15 mm working distance were used to analyze the films.

\subsection{Conductivity Measurement}
PEDOT:PSS:PEI film conductivity was measured via the two-probe method, following previously reported protocols \cite{Rauer2023}. A film was air-dried onto a microscope glass slide to immobilize the specimen. Subsequently, the film was contacted via two silver plates (25 x 25 mm, Alfa Aesar) pressed against the film using clamps to ensure a good electrical connection. The distance between the silver plates was quantified, and the width of the film was measured. The thickness of the film was determined by a digital micrometer screw gauge (FORTIS). Finally, the resistance of the film was measured with a multimeter (2025, PeakTech) and the film conductivity $\sigma$ (mS/cm) was calculated based on the following equation:

\begin{equation}
    \sigma = \frac{1}{\rho }  = \frac{L}{{R}\cdot {A}}
\end{equation}

where $\rho$ is the resistivity ($\Omega\cdot\mathrm{m}$), $L$ is the distance between measurement electrodes ($\mathrm{m}$), $R$ is the measured resistance ($\Omega$), and $A$ is the cross-sectional area orthogonal to the current flow direction (m\textsuperscript{2}).

\subsection{Tensile Strength}
A tensile testing machine (Zwick Roell) was used to study the tensile strength properties of the films with an Xforce P load cell with a maximum capacity of 100 N. The vertical tensile test measurement was carried out at 21°C and 55\% relative humidity. Specimens were tested at a strain rate of 10 mm/min. The slope of the standard force over strain inside the linear strain zone was used to calculate Young's modulus. The reported values for Young's modulus and break strain represent the mean of n = 3 replicates for each film type.

\subsection{Construction of the OECTs Cell}
An OECT cell consisting of source-, drain- and gate electrodes was placed on a glass substrate (Microscope slide, VWR). The source and drain are connected by a film channel, the PEDOT:PSS:PEI film in this work. Source, drain, and gate electrodes were conducted based on Au-coated microscope slides (layer thickness 100 Å, 99.999\% Au, Sigma-Aldrich). To immobilize the PEDOT:PSS:PEI film between the source and the drain electrode and ensure adherence to the substrate, a few drops of an adhesion PEDOT:PSS solution was applied on top of the Au pads, containing 0.05 vol\% DBSA; 1 wt\% GOPS, 5 vol\% EG and 93.95\% PEDOT:PSS dispersion \cite{Wan2015}. Subsequently, an annealing step at $140 ^\circ C$ for 90 min was conducted. As a result, a transistor channel measuring 20 mm in width (W) and 10 mm in length (L) was fabricated. An Au-coated microscope slide (10 mm × 10 mm) was used as a gate electrode.

A custom-made PDMS wall was attached to the glass substrate of each sample to contain electrolytes and form an electrochemical cell for characterization and ion-sensing measurements. A 10:1 PDMS to curing agent mixture was poured into the petri dish and cured overnight at 50 °C to manufacture the PDMS slab. The PDMS structures were subsequently removed from the petri dish while creating a rectangular prism in the middle to hold the electrolyte using a scalpel. The PDMS was then adhered to a glass substrate using two-component epoxy glue (UHU, Plus Schnellfest 5 min) for 3 h. Before gluing, the PDMS slab and the microscopy slides were cleaned with isopropanol and dried to improve the connection.

\subsection{OECTs Characterization}
All measurements were carried out in 0.1M aqueous NaCl solution as electrolyte and the 3 electrodes were immersed. The characterization of the transfer curve was performed with a 4-channel power supply (HMP4040, Rohde \& Schwarz). To characterize transient performance, the channel current I\textsubscript{D} was measured during gate voltage (V\textsubscript{G}) sweeps (from 0 to 0.6\,V in 0.1\,V steps) at multiple fixed drain voltages (V\textsubscript{D} = 0.4, 0.5, and 0.6\,V). Each constant V\textsubscript{D} condition produced a distinct I\textsubscript{D}-V\textsubscript{G} transfer curve.

For transient characteristics, the setup consisted of a 4-channel power supply connected between the gate and source electrodes and a potentiostat (PGSTAT302N, Autolab) connected between the drain and source electrodes, with the source serving as the shared ground for both instruments. The total measurement time for the experiment was 30 seconds, and the drain voltage was kept constant. At the start of the experiment, a gate voltage of 0V was applied; at 10 seconds, a gate voltage step to 0.6 V was applied, and lastly, 0V was again applied at 20 s for the remaining experiment time. To compensate for the impact of the hump during measurement, the reaction time was calculated using the amount of time required to reach 90\% of the saturation current \cite{Lee2021}.

\medskip

\textbf{Supporting Information} \par
Supporting Information is available from the authors upon request.

% Acknowledgements
\medskip
\textbf{Acknowledgements} \par
S.W. and M.R. contributed equally to this publication. This work was partly funded by the China Scholarship Council (CSC) under Grant 202108080037. M.W. acknowledges DFG funding through the Gottfried Wilhelm Leibniz Award 2019 (WE 4678/12-1). M. Wessling appreciates the support from the Alexander-von-Humboldt foundation. This work was performed in part at the Center for Chemical Polymer Technology CPT, which is supported by the EU and the federal state of North Rhine-Westphalia (grant no. EFRE 30 00 883 02). The authors thank Karin Faensen, Irma Staskiewicz, Felix Schmitz and Jingyun Xu for their support and contributions.

% References (arXiv: ship mybibfile.bib with the submission, or paste the .bbl)
\bibliographystyle{unsrt}
\bibliography{mybibfile}

\end{document}